\documentclass[sigconf]{acmart}

\usepackage{tabularx}
\usepackage{hyperref}

\AtBeginDocument{%
  \providecommand\BibTeX{{%
    \normalfont B\kern-0.5em{\scshape i\kern-0.25em b}\kern-0.8em\TeX}}}

\copyrightyear{2025}
\acmYear{2025}
\acmConference[CHI '25]{CHI Conference on Human Factors in Computing Systems}{April 26-May 1, 2025}{Yokohama, Japan}
\acmBooktitle{CHI Conference on Human Factors in Computing Systems (CHI '25), April 26-May 1, 2025, Yokohama, Japan}\acmDOI{10.1145/3706598.3713828}
\acmISBN{979-8-4007-1394-1/25/04}

\begin{document}

\title[A Foundation for the Temporal Analysis of Dark Patterns]{Getting Trapped in Amazon's ``Iliad Flow'': A Foundation for the Temporal Analysis of Dark Patterns}


\author{Colin M. Gray}
\email{comgray@iu.edu}
\orcid{0000-0002-7307-1550}
\affiliation{
  \institution{Indiana University}
  \city{Bloomington}
  \state{Indiana}
  \country{USA}
}
  
\author{Thomas Mildner}
\email{mildner@uni-bremen.de}
\orcid{0000-0002-1712-0741}
\affiliation{%
  \institution{University of Bremen}
  \country{Germany}
}

\author{Ritika Gairola}
\email{rgairola@iu.edu}
\orcid{0009-0003-5143-2807}
\affiliation{
  \institution{Indiana University}
  \city{Bloomington}
  \state{Indiana}
  \country{USA}
}

\renewcommand{\shortauthors}{Gray, Mildner, \& Gairola}

\begin{abstract}
  Dark patterns are ubiquitous in digital systems, impacting users throughout their journeys on many popular apps and websites. While substantial efforts from the research community in the last five years have led to consolidated taxonomies and an ontology of dark patterns, most characterizations of these patterns have been focused on static images or isolated pattern types. In this paper, we leverage documents from a US Federal Trade Commission complaint describing dark patterns in Amazon Prime's ``Iliad Flow,'' illustrating the interplay of dark patterns across a user journey. We use this case study to illustrate how dark patterns can be characterized and mapped over time, providing a sufficient audit trail and consistent application of dark patterns at high- and meso-level scales. We conclude by describing the groundwork for a methodology of Temporal Analysis of Dark Patterns (TADP) that allows for rigorous identification of dark patterns by researchers, regulators, and legal scholars. 
\end{abstract}



\maketitle

\section{Introduction \& Background}
After over a decade, interest in dark patterns research is growing---impacting not only HCI scholarship, but also connecting issues across policy, law, and design~\cite{Gray2023-zc}. The study of dark patterns was originally focused on design practitioners with the goal of ``naming and shaming'' companies into providing better user experiences~\cite{Brignull_undated-vj}, but has since captured a range of patterns that deceive, coerce, or manipulate users. Researchers have developed systematic knowledge of dark pattern instances in a range of domains, including e-commerce~\cite{Mathur2019-hx}, games~\cite{Zagal2013-gj}, and social media~\cite{Mildner2023-qo}. Additionally, dark patterns have been studied in many specific contexts, including extended reality (XR; \cite{Krauss2024-ar}), consent banners~\cite{Soe2020-mq}, and mobile apps~\cite{Di_Geronimo2020-mh}. These domains and contexts are further elaborated in systematic reviews of the dark patterns literature by \citet{Gray2023-zc} and \citet{Chang2024-hx}. These efforts have most recently supported the development of a domain-agnostic ontology~\cite{Gray2023-ds,Gray2024-yt}, which has categorized individual dark pattern types into high-, meso-, and low-level patterns to allow easier access and adaption within and outside this community. Following definitions proposed by \citet{Gray2024-yt}, high-level types encompass context- and domain-agnostic \textit{strategies}, meso-level types formulate a context- and domain-agnostic \textit{angle of attack} that relates to a higher level strategy but with slightly less abstraction, while low-level types contain context-specific dark patterns as \textit{applied in the user interface}. 

In parallel with research on dark patterns\footnote{We use this term to connect our efforts to prior scholarship and legal statute, while recognizing that other terms such as ``deceptive design'' or ``manipulative design'' are sometimes used to describe similar tactics. While the ACM Diversity and Inclusion Council has included dark patterns on a \hyperlink{https://www.acm.org/diversity-inclusion/words-matter}{list of potentially problematic terms}, there is no other term currently in use that describes the broad remit of dark patterns practices that include deceptive, manipulative, and coercive patterns that limit user agency and are often hidden to the user.}, regulators and policymakers have begun to take action to address deceptive design practices, resulting in legislative frameworks and regulatory sanctions that aim to protect users from dark patterns' harm. Legislation such as the EU's Digital Service Act (DSA)~\cite{DSA2022} and California's Privacy Rights Act (CPRA)~\cite{CPRA}, alongside guidance from governmental bodies such as the Organization for Economic Co-operation and Development (OECD)~\cite{Oecd2022}, UK Consumer and Markets Authority (CMA)~\cite{CMA2022}, and the US Federal Trade Commission (FTC)~\cite{FTC2022} have supported action against dark patterns, with the goal of increasing transparency and protecting users' autonomy to make informed decisions. Currently, lawsuits and other sanctions are leveraging these new regulations and, thus, demonstrate the effectiveness of policies where HCI and law work side-by-side to protect end users as envisioned by \citet{Yang2024-mu} and \citet{Jackson2014-wc}.

However, existing research often focuses on static dark patterns, driven by sharing screenshots or similar artifacts as evidence of dark patterns latent in the UI~\cite{Gray2018-or}. While some dark patterns scholarship acknowledges aspects of temporal experience~\cite{Mildner2023-qo,Gray2021-zf}, including feedforward, repetition of actions such as nagging, or actions that are part of a larger sequence of user interactions, no expert evaluation or automated methods have been proposed that comprehensively support the inspection and synthesis of dark patterns across an entire user journey. 

    



Numerous scholars have called for more attention to this temporal detail as a means of documenting the aggregate impacts of dark patterns on end users~\cite{Oecd2022,Gray2023-kq,Luguri2021-bg}. Recent work from \citet{Mildner2023-qo,Mildner2023-dd}, echoing prior work from \citet{Gray2021-zf}, \citet{Luguri2021-bg}, and guidance from the OECD~\cite{Oecd2022}, suggests that not only do dark patterns often occur together in single moments of a user journey (co-occurrence), but they can also produce amplified effects both in isolation and across a user journey. We advance this line of research in this paper, laying the foundation for a method of Temporal Analysis of Dark Patterns (TADP) by considering necessary components illustrated through a case study from the legal literature.

To that end, we make two contributions to the HCI and dark patterns literature. First, we illustrate aspects of temporal complexity that impact user experiences of dark patterns, leveraging a US Federal Trade Commission complaint regarding the Amazon Prime ``Iliad Flow'' as an explanatory case study\footnote{\textit{The Iliad} is a Greek epic by Homer which tells the story of Achilles in the long and drawn out Trojan war~\cite{homer_iliad_2023}. Amazon internally used ``Iliad'' as a project title to indicate the difficulty users faced when attempting to unsubscribe from Amazon Prime.} to characterize how dark patterns occur over multiple pages and as part of a service delivery strategy. Using this temporally-sensitive methodology, researchers, regulators, and legal scholars can build a shared understanding of how best to describe the presence and impact of dark patterns on user interactions over time. Second, we demonstrate a TADP methodology researchers can use to document how dark patterns impact user experiences, identifying relevant components to visualize as part of rigorous expert evaluation. 

\section{Levels of Temporality for Analysis of Dark Patterns}
\label{sec:levels}
A user experience unfolds over time~\cite{Karapanos2009-mi}. Designers can leverage this temporality to build deception into digital systems, using sophisticated design techniques across different stages of the user experience to manipulate, deceive, or coerce the end user. \citet{Gray2021-jv} have previously proposed a continuum that describes how users experience manipulation over time, moving from initial judgment to ultimate (negative) impacts. Building on this and other prior scholarship~\cite{Mildner2023-qo,Mildner2023-dd,Oecd2022,Luguri2021-bg}, we argue for the existence of at least three different levels of temporality through which we can analyze, characterize, and provide a visual synthesis of dark patterns. For each level, we pair our analysis with a common user experience (UX) representation method that is used to depict interactions within and across time, demonstrating how that level of analysis can be activated by researchers in our case analysis. 

\begin{itemize}
    \item \textbf{Intra-Page:} Intra-page temporality refers to the use of dark patterns on a single page, including in-page interaction, micro-interactions, or other user actions that do not cause a user to navigate to a new page. The temporal effect of these dark patterns on a single page relates to sequential and/or simultaneous occurrence and the cascade of these patterns in relation to specific user decisions that may impact the user’s decision-making (an instance of \citet{Gray2021-jv} ``initial judgment'' $\rightarrow$ ``general conclusion''). \textbf{Visualization: annotated wireframes.}
    \item \textbf{Inter-Page:} Inter-page temporality refers to the use of dark patterns across multiple pages, whereby patterns within and across pages collectively produce manipulative impacts that shape the user experience. The temporal effect of these patterns builds upon the effect of the previously experienced dark patterns by either mirroring prior dark patterns or by adding a new layer of manipulation through an additional dark pattern. The temporality of dark patterns over multiple pages can be understood as they emerge at each stage of the user journey collectively deterring the user from completing the intended task (an instance of \citet{Gray2021-jv} ``undesired interaction'' or ``negative results from interaction''). \textbf{Visualization: task flows.}
    \item \textbf{System:} System-level temporality refers to the use of dark patterns in the overarching architecture and design of the digital service. The temporal effects of these patterns are often hidden to the end user, and are either experienced collectively across multiple touchpoints or action is forced at the code or service delivery level (an extension of \citet{Gray2021-jv} ``negative results from interaction''). \textbf{Visualization: service blueprint. }
\end{itemize}

Overall, we argue that these three levels operate together in impacting the user experience, and at each level, dark patterns may be experienced by users in different ways that lead to a collective experience of deception, manipulation, and/or coercion. Next, we illustrate these levels of temporality as an analytic tool through a description of a contemporary case of dark patterns enforcement.

\section{Problematizing Dark Patterns Experienced Over Time: A Case Study of Amazon Prime's ``Iliad Flow''}
In order to explain how the consideration of temporality advances the study of dark patterns, we include an explanatory case study~\cite{Yin2009-vs} based on a common instance of dark patterns---account deletion or unsubscribing from a service---which has previously been shown to be a challenge for users and is often linked with economic harms~\cite{Schaffner2022-nn,Sheil2024-xu,Kaldestad2021-cj}. This context is particularly timely, given a recent rule enacted by the US FTC known as  \textit{Click to Cancel}, which will require a subscription to be as easy to cancel as it was to sign up~\cite{FTC_2024-ld} and is similar in function to other EU sanctions regarding service cancellation~\cite{Kaldestad2021-cj}. The case we describe builds upon a legal complaint filed by the US FTC in June 2023 against Amazon which includes detailed references to dark patterns~\cite{FTCAmazon2023-sr}\footnote{Numerous other legal cases invoking the term have been catalogued at \url{https://www.deceptive.design/cases}.}. This case follows numerous others~\cite{EpicGames2022,ftc-2022-lending,ftc-2021-abcmouse} 
that have used the presence of dark patterns as a central form of evidence that user autonomy was not respected. With our aim to aid transdisciplinary action across HCI and law~\cite{Gray2021-zf,Gray2023-kq}, we leverage this legal complaint to identify which dark patterns are inscribed into the user experience, how these dark patterns relate to each other on specific pages and over time, and what elements of the overall user experience would be useful for scholars to focus on when analyzing other experiences for the presence of dark patterns. Importantly, we do not focus our contribution on the process of detecting dark patterns in isolation (well addressed by prior work; e.g.,~\cite{Mathur2019-hx,Mathur2021-rc,Mildner2023-dd}), but rather use the case to illustrate where new visualization and synthesis methods can enhance descriptions of temporal complexity in contemporary legal proceedings. By doing so, we advance a novel methodology for temporal analysis that supports key claims within the case while also subsuming other legal analyses regarding the presence of dark patterns. 

A Norwegian Consumer Council report from 2021 previously demonstrated how dark patterns in Amazon's cancellation process frustrated consumers, leaving them ``[\ldots] faced with a large number of hurdles, including complicated navigation menus, skewed wording, confusing choices, and repeated nudging. Throughout the process, Amazon manipulates users through wording and graphic design, making the process needlessly difficult and frustrating to understand.''~\cite{Kaldestad2021-cj}. In the FTC complaint against Amazon~\cite{FTCAmazon2023-sr}, these same allegations were expanded, showing how Amazon's design teams were complicit in making this process more difficult than it needed to be:

\begin{quote}
  \textit{[\ldots] the primary purpose of the Prime cancellation process was not to enable subscribers to cancel, but rather to thwart them. Fittingly, Amazon named that process ``Iliad,'' which refers to Homer’s epic about the long, arduous Trojan War. Amazon designed the Iliad cancellation process (``Iliad Flow'') to be labyrinthine, and Amazon and its leadership [\ldots] slowed or rejected user experience changes that would have made Iliad simpler for consumers because those changes adversely affected Amazon’s bottom line.} \cite[p. 3]{FTCAmazon2023-sr}
\end{quote}

The complaint features an exhaustive description of the user journey, supported by screenshots. Several different aspects of the interactive system are included in the complaint, including the process to subscribe to Amazon Prime, different ways to enter the ``Iliad Flow'' to cancel Amazon Prime, and the component interactions required to cancel the membership.

\subsection{Method for Identifying Dark Patterns}
The FTC complaint~\cite{FTCAmazon2023-sr} includes explicit analysis that identified multiple dark patterns across the ``Iliad Flow.'' To characterize these instances of dark patterns in more detail, we carefully evaluated the complaint text, identified and reconstructed screenshots of the Amazon interfaces, and noted examples of named dark patterns identified by the FTC. All researchers reviewed the complaint text to perform these tasks, 
leveraging our research team's experience in UX design, including methods such as wireframes, task flows, user journey maps, and service blueprints. We also benefited from our research team's deep expertise in qualitative content analysis---the most common method used to assess the presence of dark patterns~\cite{Gray2023-zc}. All three members of the research team had legal experience, one as a lawyer (Ritika), one who has previously written expert reports relating to analysis of dark patterns in court cases utilizing similar methods (Colin), and one who has contributed to expert reports relating to dark patterns (Thomas).

Using evidence from the complaint, we reconstructed key portions of the user experience,
\footnote{Legal proceedings typically rely upon documents disclosed by companies through a formal discovery process. While HCI scholars often use videos to perform temporal analysis, in this case, only screenshots were disclosed in the FTC complaint, so we had to reconstruct the user experience and task flows with those materials.} organizing screenshots into task flows, identifying where multiple screenshots indicated how a user might interact with the system, and connected multiple activities performed over a period of time. We then documented instances of dark patterns in these snapshots of the user experience, using the \citet{Gray2023-ds,Gray2024-yt} ontology vocabulary of high, meso, and low-level patterns. Where possible, we utilized identification of dark pattern types already identified in the complaints, mapping them to the relevant pattern type and level(s) in the ontology. However, by using a qualitative content analysis approach~\cite{Mildner2023-qo}, we also identified additional instances of dark patterns. Our analysis facilitated flexible investigation of interface components that can be both static and experienced over time. One researcher performed the initial identification of dark patterns from the complaint, and all three researchers then confirmed the presence and mapping of these dark patterns to the ontology based on their prior experience using qualitative content analysis to identify the presence of dark patterns. 
We then connected the different interface stages in the form of a task flow, which enabled us to identify dark patterns that were used in conjunction with each other (co-occurrence), dark patterns which repeated over multiple pages or points of user interaction, and instances where the co-occurrence or persistent use of one or more dark patterns resulted in the potential amplification of effect on the user. 

In the following sub-sections, we report on the findings of this analysis at the three different levels of temporality previously discussed in Section~\ref{sec:levels}: intra-page, inter-page, and system-level.

\subsection{Intra-Page Dark Patterns}
\label{sec:intrapage}
In Figure~\ref{fig:intrapage}, we show two individual pages that are part of the ``Iliad Flow,'' each includes annotations that demonstrate how static dark patterns are used within the user interface. The left page depicts the page the user is shown once they successfully navigate the help system and discover the beginning of the Iliad Flow (see more about this discovery process in Section~\ref{sec:system}). The right page depicts the second page of the Iliad Flow. We start our analysis by demonstrating how dark patterns exist in isolation on each page:

\begin{figure*}[t!]
    \centering
    \includegraphics[width=0.95\textwidth]{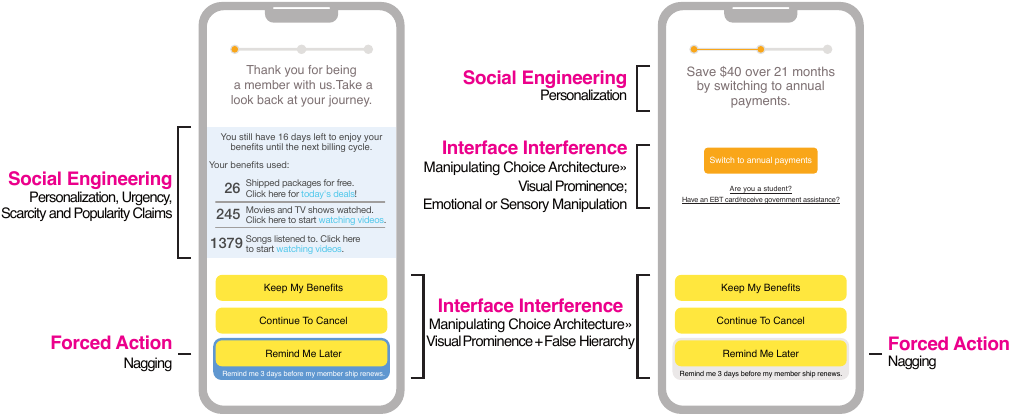}
    \caption{An annotated wireframe of two of the pages in the Iliad Flow. Each annotation illustrates specific instances of high-level strategies (in \textbf{bold}) and meso-level dark patterns in the user interface. This figure includes reconstructed screenshots from the FTC complaint~\cite{FTCAmazon2023-sr}.}
    \Description{An annotated wireframe of two of the pages in the Iliad Flow. Two different pages are represented with a range of interface elements that have been annotated with specific types of dark patterns, including high-level strategies (in \textbf{bold}) and meso-level dark patterns in the user interface. This figure includes reconstructed screenshots from the FTC complaint~\cite{FTCAmazon2023-sr}.}
    \label{fig:intrapage}
\end{figure*}

\begin{itemize}
    \item On both pages, there is evidence of the high-level dark pattern \textbf{Social Engineering}. On the left page, social engineering is integrated through the meso-level pattern of \textit{personalization}, using detailed statistics to show a user about their past activity (presumably related to Amazon Prime benefits). On the right page, personalization dark patterns are also used to show potential savings related to shifting from monthly to annual payments. The left page also incorporates the meso-level patterns \textit{urgency} and \textit{scarcity} and popularity claims, using common benefit categories as proof of the utility of the service’s popularity juxtaposed with the admonishment ``You still have 16 days left to enjoy your benefits until the next billing cycle.'' (emphasis in original).
    \item Both pages also contain evidence of the high-level dark pattern \textbf{Interface Interference}. On each page, the choice architecture is manipulated by including meso-level patterns \textit{false hierarchy} and \textit{visual prominence}. On each page, the first choice is meant to prevent the user from cancelling the service (``Keep My Benefits'' on the left and ``Keep My Membership'' on the right) and the third choice is meant to remind the user later. Only the second option of ``Continue to Cancel'' allows a user to continue forward in the Iliad Flow. Notably, visual prominence is used to highlight the ``Remind Me Later'' option, with a larger target size, an annotation indicating to the user that they will be reminded three days before the service renews, and stronger visual weight due to the bolder color on the first page. On the right page, emotional or sensory manipulation is also used to target users who might have less financial resources (e.g., students, those who receive government assistance), using cues to get them to stay enrolled in the service. 
    \item Finally, both pages include reminder buttons as part of the primary choice architecture, which, as a recurring feature, function as part of the \textbf{Forced Action} meso-level pattern \textit{nagging}.
\end{itemize} 

On a single page, these patterns can emerge in a sequence, building upon the preceding dark pattern’s manipulation. They also can emerge collectively, thereby amplifying the manipulative effect. The use of interface interference and social engineering dark patterns introduces unnecessary friction delaying the user’s ability to cancel the membership or nudge them towards alternative paths. 

\subsection{Inter-Page Dark Patterns}
\label{sec:interpage}
\begin{figure*}[t!]
    \centering
    \includegraphics[width=.95\textwidth]{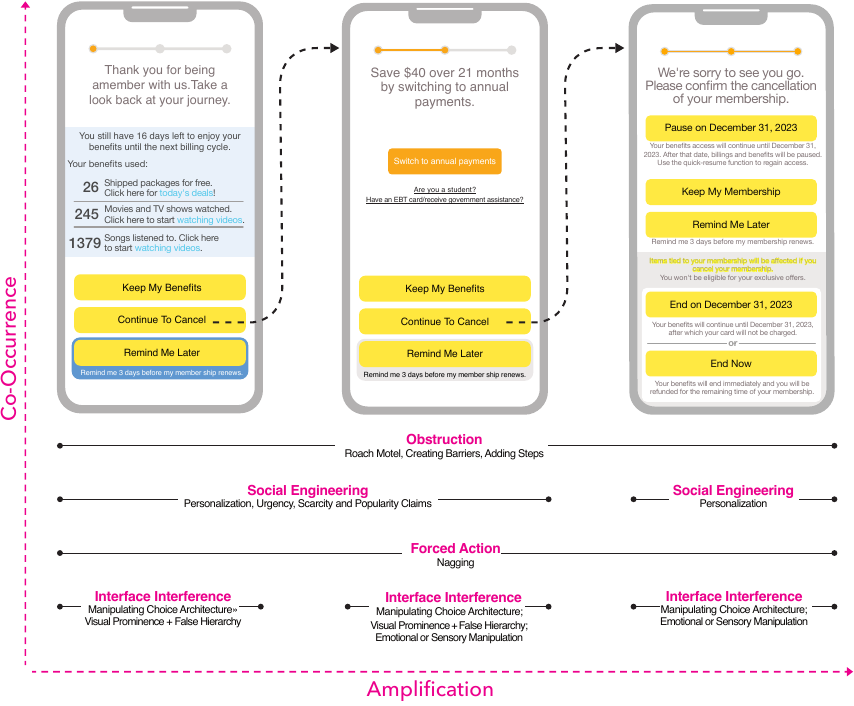}
    \caption{A task flow indicating how three different pages of the Iliad Flow are related to one another and the overall user experience. Dotted lines with arrows indicate how a user navigates from one page to the next. Lines at the bottom indicate where dark patterns are present in the user experience over time, including both high-level strategies (in \textbf{bold}) and meso-level patterns. This figure includes reconstructed screenshots from the FTC complaint~\cite{FTCAmazon2023-sr}.}
    \Description{An annotated wireframe of three connected pages representing the Iliad Flow. Each page is connected to the next through a dotted line with an arrow, indicating how a user would navigate between pages. A dotted line from bottom to top is labeled as "co-occurrence" while a line with arrow from left to right is labeled as "amplification." Below the pages, a range of dark pattern types are indicated, showing the degree to which that pattern type is present across all pages or is only present on a single page. This figure includes reconstructed screenshots from the FTC complaint~\cite{FTCAmazon2023-sr}.}
    \label{fig:interpage}
\end{figure*}

The process to cancel the Amazon Prime membership extends over five pages, which are filled with options that are both repetitive and confusing (as shown in our intra-page analysis). In Figure~\ref{fig:interpage}, we show a portion of this user journey, illustrating the three main pages of the Iliad Flow along with connecting arrows demonstrating how a user moves from one page to the next. Importantly, if a user selects the ``Keep My Benefits'' or ``Remind Me Later'' options, the flow is disrupted, and a user has to start from the beginning.

In this analysis, we focus less on annotating the manifestation of dark patterns on a single page. Instead, we indicate through bars at the bottom of the figure which dark patterns are present over time, and for what duration (a single page, two, or more pages):

\begin{itemize}
    \item Across the task flow, the high-level dark pattern \textbf{Obstruction} is most evident, including the meso-level patterns \textit{roach motel}, \textit{creating barriers}, and \textit{adding steps}. Obstruction extends across all three pages, reified through an interface that makes something easy to get into difficult to get out of (\textit{roach motel}) by adding more steps than necessary and erecting unnecessary barriers that are not aligned with the user’s goal of cancellation.
    \item Supporting the overarching goals of \textbf{Obstruction}, three other high-level dark patterns amplify the shareholder’s goal of retaining the user instead of supporting the user’s goal to cancel the service. 
    \begin{itemize}
        \item \textbf{Social Engineering} tactics are present across the task flow, with \textit{personalization} dark patterns consistently in use across all pages to provide customized information that might influence the user’s choice. Additionally, \textit{urgency} and \textit{scarcity and popularity claims} are used in the first two pages of the task flow to convince the user of the potential value of the service based on cherry-picked data points, manipulating the information flow so that a user is less likely to proceed to the next step.
        \item \textbf{Forced Action} is used across the entire task flow through the repeated use of \textit{nagging}. This causes users to potentially second-guess their decision. Additionally, if a user inadvertently clicks on the “Remind Me Later” prompt, they exit the Iliad Flow and have to start again if they wish to continue with their initial goal of cancellation.
        \item Different forms of \textbf{Interface Interference} are used in each portion of the task flow, but all pages contain evidence of \textit{manipulation of the choice architecture} and two pages make use of \textit{emotional or sensory manipulation}. The most consistent dark pattern is the use of \textit{false hierarchy} and \textit{visual prominence} in the choice architecture, making it more likely for a user to become confused by the array of options or accidentally make a choice which is not aligned with their goal of cancellation.
    \end{itemize}
\end{itemize}

Across all of these pages within the task flow, a user’s vulnerabilities are exploited through emotional manipulation, with messages and choice architectures framed to dissuade them from their decision to cancel. The combination of dark patterns experienced within and across these pages amplifies the manipulation and reinforces potential user hesitation, and despite choosing the explicit ``Continue to Cancel'' option at each stage repeatedly, the user has to navigate through a maze of alternatives of either being reminded or pausing the membership.

\subsection{System-Level Dark Patterns}
\label{sec:system}
The Iliad Flow exists within a broader and more complex service context that encourages users to sign up with very little friction while demanding extraordinary amounts of friction to reverse that decision. At the heart of this strategy is the seamless recruitment of users into the Prime Membership, integrated into many other Amazon services such as Amazon Music, Amazon Prime Video, or making purchases on the Amazon website, both on mobile and web. In Figure~\ref{fig:system}, we present an adapted service blueprint. Service blueprinting is a common service design mapping technique that describes the interplay of multiple touchpoints over time from a user's perspective~\cite{Gibbons2017-pm}. This technique includes both front-stage (consumer-facing) and back-stage (employee or internally-facing) components of a service delivery strategy---in this case, a detailing of the interactions and processes relating to Prime subscription and unsubscription elements included within Amazon's larger ecosystem. Our service blueprint documents three customer actions relating to the subscription and cancellation process, including relevant types of dark patterns, used to steer a customer's actions. Importantly, this blueprint reveals that dark patterns persist in different forms across multiple touchpoints, such as when certain dark patterns like \textbf{Obstruction} are carried out across subscription elements or when friction is very low for signup and very high for unsubscription. Thus, the system-level analysis reveals how dark patterns are used as part of a service delivery strategy rather than isolated to a single touchpoint. We begin our analysis by identifying how a service's ecosystem can steer customers to sign up before entrapping them through obstructive dark patterns:



\begin{figure*}[t!]
    \centering
    \includegraphics[width=\textwidth]{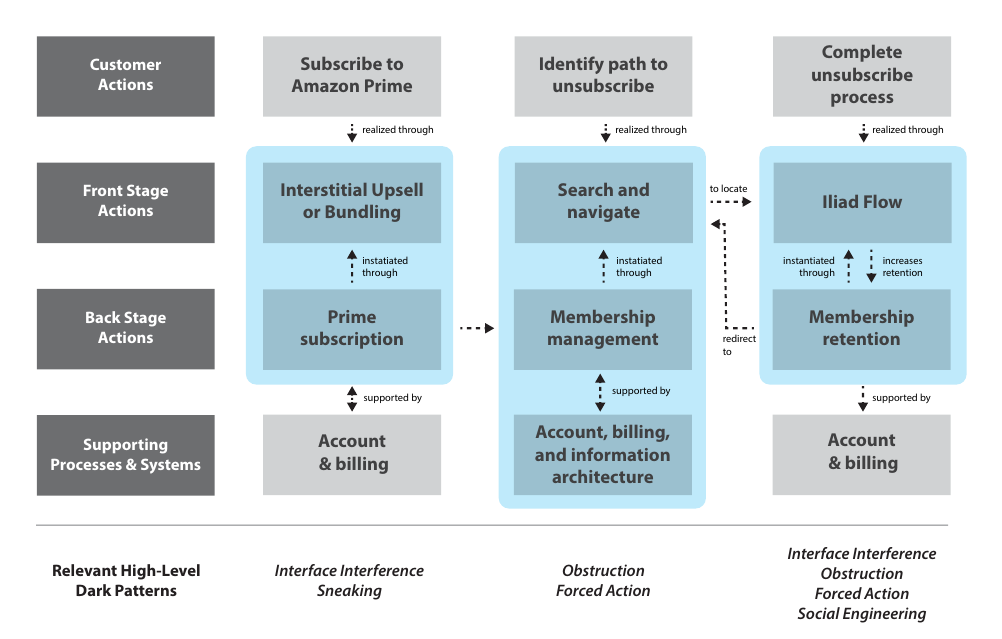}
    \caption{A service blueprint outlining some of the actions related to Amazon Prime subscriptions as part of Amazon's larger ecosystem of products and services. Customer Actions (shown at the top) indicate goals that users might have, while Front Stage Actions (second row from the top) indicate how a user might engage with the system to support those goals. Back Stage Actions and Supporting Processes and Systems (bottom two rows) indicate how Amazon would support these goals on the back end in ways that are not visible to the end user. Dark patterns are indicated in shaded portions of the blueprint with specific high-level strategies in bold italics.}
    \Description{A service blueprint outlining some of the actions related to Amazon Prime subscriptions as part of Amazon's larger ecosystem of products and services. Customer Actions (shown at the top) indicate goals that users might have, while Front Stage Actions (second row from the top) indicate how a user might engage with the system to support those goals. Back Stage Actions and Supporting Processes and Systems (bottom two rows) indicate how Amazon would support these goals on the back end in ways that are not visible to the end user. Dark patterns are indicated in shaded portions of the blueprint with specific high-level strategies in bold italics.}
    \label{fig:system}
\end{figure*}

\begin{itemize}
    \item To recruit members for their services, Amazon uses dark patterns to trick potential customers into signing up. While additional options include stand-alone and cheaper alternatives part of Amazon's wider service ecosystem (e.g., Amazon Music or Video), the \textbf{Interface Interference} dark pattern, including meso-level patterns \textit{manipulating choice architecture}, \textit{emotional and sensory manipulation}, and \textit{hidden information}, is used to promote a premium membership as a superior subscription, including all premium features but at a higher cost. Co-occurring \textbf{Sneaking} patterns like \textit{Bait and Switch} further increase the likeliness that users are being tricked into more expensive subscriptions. 
    \item Should a customer decide to leave the service or change their subscription to a stand-alone alternative, Amazon uses \textbf{Obstruction} dark patterns to make the process more cumbersome or decrease accessibility. During a customer's search for ways to unsubscribe, the service has multiple options to limit the discoverability or accessibility of cancellation options. \textit{Roach motel} meso-level patterns effectively obstruct users from unsubscribing, entrapping them in a membership plan. Co-occurring \textbf{Forced Action} patterns like \textbf{forced continuation} limit customers' choices, causing them to invest additional labor to discover the correct search criteria to locate the unsubscription options.
    \item Once the user discovers the unsubscription page, additional instances of \textbf{Obstruction} amplify previous difficulties, as details in the intra- and inter-page analyses of the Iliad Flow previously described. 
\end{itemize}

As shown in this service blueprint, Amazon deploys dark patterns strategies across its service ecosystem to steer potential customers into easily subscribing, while deploying strategies to prevent users from cancelling their membership or even discovering subscription management options. Throughout their experience with Amazon, a customers' agency is restricted through dark patterns that occur together or amplify each other, burdening the user through restricted choices and resulting in the potential for financial harm. 

\subsection{Characterizing Temporal Complexity through Co-Occurrence and Amplification}

In this section, we synthesize some findings from our temporal analysis process. 
Across our three levels of the analysis presented, our temporal mapping of the ``Iliad Flow'' and its location in the broader Amazon ecosystem revealed a plethora of dark patterns customers encounter throughout their attempt to cancel or resist an Amazon Prime membership that are strengthened by co-occurrence and amplification. 



Our analysis visually depicts how multiple dark pattern types often occur together across multiple temporal levels. As shown in Figure~\ref{fig:interpage}, high-level patterns of \textbf{Forced Action} and \textbf{Obstruction} pervaded the entire interaction sequence, supported by \textbf{Social Engineering} and \textbf{Interface Interference} in key decision moments across the three pages. Additionally, these high-level patterns were supported---and even amplified---by numerous lower-level patterns that drew on the higher-level parent types. For instance, all three pages used \textit{manipulation of the visual hierarchy} to confuse users about the interactive differences and feedforward between the three options. 
In parallel, \textbf{Social Engineering} meso-level patterns such as \textit{personalization} were used to amplify the \textbf{Interface Interference} effects by providing specific amounts of media the user might lose access to or providing options on a different payment plan that would appear more affordable. Notably, while some patterns are easily traceable to one or more specific UI elements, the interactions among the different types of dark patterns are more nuanced. For instance, the one page of the Iliad Flow layers choice architecture manipulation and emotional manipulation and \textit{urgency} in a direct way, leaving the \textit{roach motel} (\textbf{Sneaking}) to be realized across the entire user journey. Similarly, the use of \textit{adding steps} applies to the entire user journey as opposed to one discrete UI element or page.



\section{Foundations for a Temporal Analysis of Dark Patterns (TADP) Methodology}

\begin{figure*}[t!]
    \centering
    \includegraphics[width=\textwidth]{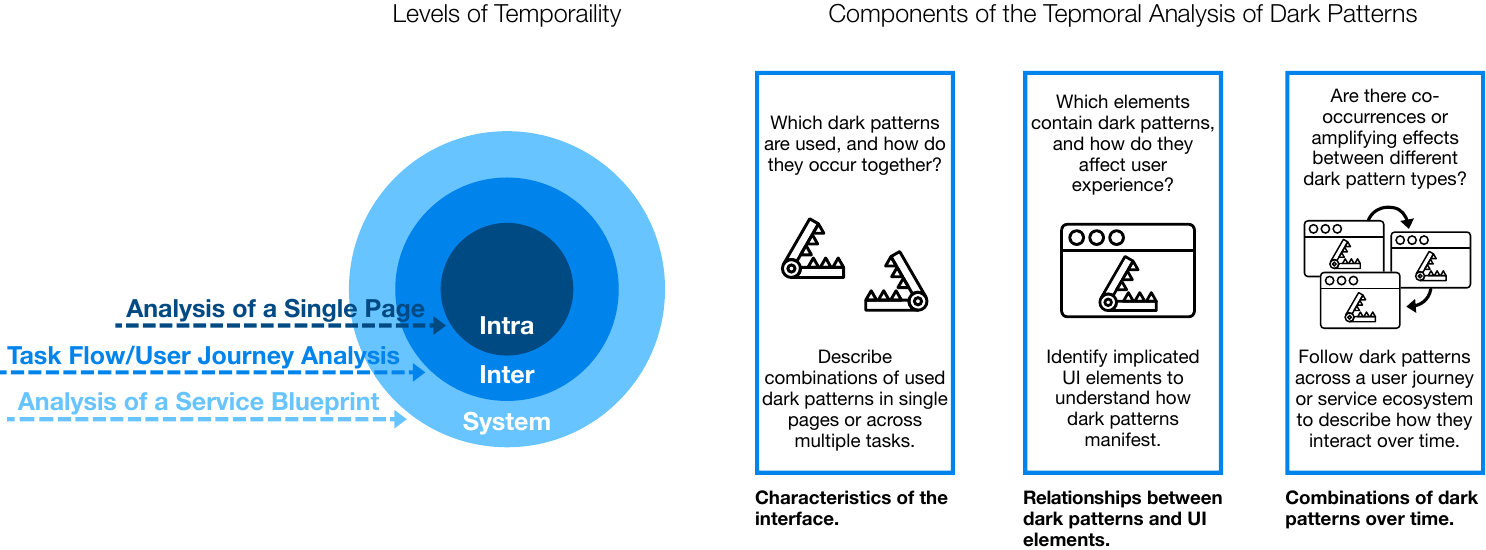}
    \caption{This figure shows the levels of temporality and components for the Temporal Analysis of Dark Pattern Methodology.} 
    \Description[Levels of temporality and components for Temporal Analysis of Dark Pattern Methodology.]{This figure follows the temporal analysis of dark patterns methodology in its foundational three components, situated next to the levels of temporality: intra, inter, and system-level analysis. A circular shape indicates the change of perspective between the levels, from analysis of a single page to task flow/user journey, and finally, the analysis of the service blueprint of a service provider. The first raises the question: Which dark patterns are used, and how do they occur together? The second asks: Which elements contain dark patterns, and how do they affect user experience? The third and last step contains the question: Are there co-occurrence or amplifying effects between different dark pattern types?}
    \label{fig:schematic}
\end{figure*}

Building on this case study, we outline key components of the Temporal Analysis of Dark Patterns methodology, along with how these components can be supported by expert evaluation. Figure~\ref{fig:schematic} offers a summary of individual steps while considering intra, inter, and system levels of temporality in which dark patterns can occur. All of these steps are recursive and iterative, recognizing that analysis on a system level might yield new findings of dark patterns that may not be immediately visible at the intra- or inter-page levels, and vice versa.

\begin{enumerate}
    \item \textbf{Identify which dark patterns are being used, in what combination or sequence, and of what type(s).} This stage requires the use of a standardized source of pattern types and definitions, such as the ontology of dark patterns by Gray and colleagues~\cite{Gray2023-ds,Gray2024-yt}. Identification of dark patterns includes high- and meso-level characterization where possible, since novel dark patterns can only be characterized 
    by an existing low-level pattern. This stage of analysis takes into account the following UI elements or characteristics: readable text; layout; relative size and positioning; use of color, typography, or text-decoration; feedforward or other forms of user feedback; task flows or other relations between UI elements and pages; and the context or medium of use. This identification of a combination of UI elements and dark pattern types indicates how the interface is manipulating, coercing, or deceiving the user, potentially using mappings to Mathur et al.'s~\cite{Mathur2021-rc} dark patterns attributes that relate to modification of the decision space or manipulation of the information flow to illustrate the mechanism(s) the dark pattern is relying upon. This identification process clearly indicates which dark patterns are experienced in a static way (e.g., visual prominence of a button), which are experienced through a user's interaction with a single page (e.g., scrolling or micro-interactions), and which are experienced cumulatively over two or more pages of the user journey (e.g., obstruction that makes an action unnecessarily difficult).
    \item \textbf{Identify which interface element(s) are implicated in the use of dark patterns, and how these concentrations of elements within the interface might lead to the user's experience of dark patterns.} This stage requires connections between the presence of a dark pattern and its manifestation in the interface or system. This stage of analysis takes into account the relationship between: one or more dark patterns to one or more interface elements; one or more dark patterns to the lack of visible interface elements; or one or more dark patterns to transitions between pages or across the entire user journey. Different levels of dark pattern characterization allow characterization of high- and meso-level patterns on the pages or journey level that are then inscribed into one or more specific interface elements. Building on the identification of dark patterns that are statically experienced, experienced through interaction on a single page, and/or experienced across two or more pages, this identification of interface elements provides a temporal account of how a user is likely to interpret and respond to the interface elements and the potential downstream impacts of this interaction.
    \item \textbf{Describe interactions between dark patterns, co-occurrence of dark pattern types, and/or potential amplification effects.} This stage requires knowledge of which dark pattern types appear and in which combination, both on a specific page and over time. This stage of analysis takes into account the: combinations of dark patterns that appear in discrete moments of the user journey and over time; the co-occurrence of patterns with shared or differing high- or meso-level parents; the strategies or cognitive biases the patterns exploit; and the causal or other interactive relationship between patterns on a page or over time. The net result of these interactions in relation to the user experience is considered in this description, demonstrating both the impacts of individual moments of the user journey where multiple dark patterns co-occur and also the cumulative impacts of dark patterns that are experienced in different combinations and sequences over multiple steps of a user journey. This description can then be evaluated for amplification effects or combinations of dark patterns that may adversely or disproportionately impact vulnerable user groups.
\end{enumerate}

We anticipate that this foundational methodology for Temporal Analysis of Dark Patterns (TADP) will advance the expert evaluation techniques used to identify and characterize dark patterns, adding explicit attention to the temporality of user experience and contributing concrete visualization approaches to synthesize the temporal experience of dark patterns. This methodology provides a robust evidence-gathering approach to both identify and characterize dark patterns in interactive systems---the first comprehensive approach that spans from single pages to an overarching service delivery strategy---facilitating disciplined analysis of digital systems to support legal proceedings and academic research. This methodology requires consideration of researcher skill, positionality, and evidence-gathering approaches that warrant further conversation. Because automated or semi-automated techniques to identify dark patterns are still relatively nascent (e.g.,~\cite{Bouhoula-2023,Koch-2023,Mathur2019-hx,Jieshan2024-ga}), our focus in this paper describes TADP as an expert evaluation method that requires a transparent and rich audit trail to support both legal and academic research goals. In order to achieve this rigor, expert evaluation relies on the \textit{expertise} of the researcher conducting the analysis---which in this methodology includes: 1) knowledge of UX representational forms (e.g., wireframes, task flows, user journey maps, service blueprints); 2) the ability to intepret the interactive qualities of user interface elements and the ways in which these UI elements contribute to the larger user experience (cf.,~\cite{Gray2021-zf,Bardzell2011-hv,Janlert2017-vi}); and 3) knowledge of common dark patterns, their manifestations, definitions, and potential harms as it relates to the user experience. 

Importantly, while the TADP methodology may excel at identifying specific combinations of dark patterns in particular user interfaces, because of the need for relatively extensive reconstructive analysis, this methodology may not be ideal for ``sweeps'' assessing dark patterns at larger scales (e.g.,~\cite{ICPEN-ch}). Additionally, since this methodology requires access to artifacts that allow for reconstruction of the user experience (e.g., screenshots, indications of task flow or feedforward), evidence that only exists as static screenshots may be difficult to assess without additional information on how a user may interact with the UI elements over time.

\section{Conclusion}
In this short paper, we present a case study of the Amazon Prime ``Iliad Flow'' to characterize the complexity of dark patterns as they are experienced over time. We used this case to demonstrate how dark patterns exist in combination and over time across three different levels of temporality, supporting the foundation for an analysis methodology for Temporal Analysis of Dark Patterns (TADP). We identify key components of this methodology and describe how expert evaluators can identify, characterize, and visually synthesize the presence of dark patterns in compelling ways that support regulation, legal proceedings, and scholarship. 

\begin{acks}
 We gratefully acknowledge feedback from Nataliia Bielova, Cristiana Santos, and other participants from the Lorentz Center ``Fair Patterns for Online Interfaces'' workshop that improved our temporal analysis methodology. This work is funded in part by the National Science Foundation under Grant No. 1909714 and the Leibniz ScienceCampus Bremen Digital Public Health, which is jointly funded by the Leibniz Association (W72/2022), the Federal State of Bremen, and the Leibniz Institute for Prevention Research and Epidemiology – BIPS.
\end{acks}
\bibliographystyle{ACM-Reference-Format}
\bibliography{temporalanalysis}

\end{document}